\documentclass[12pt,a4paper,doublespace]{amsart}

\usepackage{subfigure,graphicx,graphics}
\usepackage{amsmath,amsfonts,latexsym,amssymb,euscript,xr,url}

\usepackage{algorithmic}
\usepackage{algorithm}
\usepackage{url}
\usepackage{subfig}

\title{The impact of Gene Ontology evolution on GO-Term Information Content}

\author{Pietro Hiram Guzzi $^{(1)}$, Giuseppe Agapito$^{(1)}$,   Marianna Milano $^{(1)}$, Mario Cannataro $^{(1,2)}$,}
\address{(1) Department of Medical and Surgical Sciences\\
University of Catanzaro, Italy, \{agapito,m.milano,cannataro,hguzzi\}@unicz.it
\\ 
(2)ICAR-CNR\\ Rende, Italy, \
}

\begin{document}
\maketitle
%\doublespace
\thispagestyle{myheadings}
\pagestyle{myheadings}
%\markright{\tt Draft}%check year

\begin{abstract}
The Gene Ontology (GO) is a major bioinformatics ontology that provides structured controlled vocabularies to classify gene and proteins function and role. The GO and its annotations to gene products are now an integral part of functional analysis. Recently, the evaluation of similarity among gene products starting from their annotations (also referred to as semantic similarities) has become an increasing area in bioinformatics. While many research on updates to the structure of GO and on the annotation corpora have been made, the impact of GO evolution on semantic similarities is quite unobserved.  Here we extensively analyze how GO changes that should be carefully considered by all users of semantic similarities. GO changes in particular have a big impact on information content (IC) of GO terms. Since many semantic similarities rely on calculation of IC it is obvious that the study of these changes should be deeply investigated.  Here we consider GO versions from 2005 to 2014 and we calculate IC of all GO Terms considering five different formulation. Then we compare these results. Analysis confirm that there exists a statistically significant difference among different calculation on the same version of the ontology (and this is quite obvious) and there exists a statistically difference among the results obtained with different GO version on the same IC formula. Results evidence there exist a remarkable bias due to the GO evolution that has not been considered so far. Possible future works should keep into account this consideration.
\end{abstract}

%\doublespacing
\section{\bf Background}

Ontologies are more and more used in bioinformatics and computational biology since they provide a structured and uniform vocabulary of terms useful to describe a domain \cite{Couto:SEMANTICSIMILARITIBIOMEDONTO:PLOS2009}.

For instance, the Gene Ontology (GO) \cite{Harris:GeneOntology:NAR2004} is a large vocabulary of terms (namely GO Terms) containing information about gene products. It is structured on three main taxonomies: biological processes (BP), molecular functions (MF) and cellular components (CC). Each taxonomy is modeled as a directed acyclic graph (DAG), with the edges representing relationships between the categories. There also exist non-taxonomical relations representing relations of \textit{regulates}, \textbf{has\_part} that have been recently introduced. Considering \textit{is\_a} relationships, it should be noted that  higher-level terms represent more general descriptions. 

GO Terms are mainly used to formally describe genes and gene products\cite{GOReferenceGenome}, of different species. Moreover, the set of annotations available for each genome (or proteome) has been quickly used in different application of analysis, demonstrating its usefulness and relevance \cite{duPlessis}. 

% Frase su functional enrichmnet
For instance, a classic application of GO is called \textit{functional enrichment analysis}, i.e. given a set of gene products functional enrichment algorithms aim to identify a sub set of GO terms that are significantly more present than expected at random \cite{Huang01012009}. Other application of GO are represented, for instance, in the use of GO to model bioinformatics application or to guide the composition of workflows \cite{Cannataro2010}.

% frase su per definire semantic similarity

More recently, the quantification of  similarity among terms belonging to GO by using Semantic Similarity Measures (SSM) has gained an important place. A SSM takes in input two or more terms of GO and produces as output a numeric value in the $[0..1]$ interval representing their similarity. Since gene products are annotated with set of GO  the use of SSMs to evaluate the functional similarity among gene products is becoming a common task. Consequently the use of SSMs to analyze biological data is  gaining a broad interest from researchers \cite{Couto2007137,Guzzi2012}. Nevertheless, the use of SSMs presents some drawbacks and some limitations due to the structure of each SSM \cite{Guzzi2012}. SSM are based on two main pillars: (i) the structure of GO, and (ii) the annotation corpora (i.e. the set of available annotation for each specie). In particular many SSMs rely on different schemas of calculation of Information Content of GO terms \cite{Guzzi2012}. There exist two ways to define IC: extrinsic IC calculation  involves annotation data for an considered corpus, while intrinsic IC is based on structural information extracted from  the GO DAG \cite{harispe2013framework}. %In this way, it relies only on the intrinsic topology of the GO structure and  the dependence on annotated corpora is prevented avoiding data circularity problems.  Intrinsic IC calculus can be estimated using different topological characteristics as ancestors, number of children, depth (for lack of space we do not discuss all the other methods, see \cite{harispe2013framework} for a complete review)

SSM are thus based on GO and its annotations, but ongoing scientific research causes the regular update of both GO and its annotation. 

In particular  GO changes are regularly made available by the GO Consortium \cite{GOConsortium,GOChanges:HUNTLEY:2013} causing modification both on structure of GO and on quantity and quality of annotations. Many changes are made to both the ontology and annotation sets over time - some of these changes are planned and announced by GOC or its members via mailing lists or release notes. Other changes are not planned and made on the basis of needed  improvements, such as user requests for updates to the ontology or annotations, as well as quality assurance checks. For instance, as reported by \cite{hartung_dils2008}, from 2005 to 2008 there have been more than 15.000 changes in GO structure comprising both addition and deletion of terms \footnote{\url{http://dbs.uni-leipzig.de/research/projects/bioinformatik/ontology_evolution/go_biological_process}}. 

The evolution of GO has been analyzed deeply in the past \cite{complexityGOEvolution}. Actually there also exist tools for direct analysis of GO evolution, such as CODEX \cite{HartungOndex}. Typical modification of GO are the inclusion of novel GO Terms, the modification of the GO Tree structure and the deletion of GO terms. Modification may involve both  leaves and intermediate nodes. These ontological modifications may cause changes in the annotations e.g. when a category is removed, the annotations need to be moved or deleted. Analogously some  annotations may be modified to reflect new discoveries or to eliminate inconsistencies \cite{dessimozquality}.

Each taxonomy in GO evolved at different rates and in different ways. For instance, as noted in  \cite{HartungOndex} between 2007 and 2010, BP increased by about 70\%, compared with CC (40\%) and MF (20\%). The impact of changes in GO structure has been studied on Functional Enrichment algorithms \cite{Gross15102012} has been studied, while few studies have been investigated the impact on SSMs.

We here focus on this consideration, looking for example, how these changes may lead to different results in SSM. In particular we here analyze the impact of ontology changes on calculation of IC aiming at the determination of the impact of these changes on GO and the presence of potential more relevant changes on the structure. For example, additions of categories at the leaf level might be less critical than structural revisions within the ontology.

We considered GO version since 2005 to 2014 and formulation of IC provided by Zhou et al  \cite{zhou2008new}, Sanchez \cite{sanchez2011ontology}, Sanchez and Harispe \cite{Harispe2013}. For each version of GO we calculated the IC of all the GO terms. We then performed two kind of tests aiming to discover two possible causes of bias: (i) \textit{within IC} aiming to discover possible difference of the IC distribution dependent from the GO version, (ii) \textit{between IC} aiming to discover possible differences of GO distribtion considering the same GO version and different IC. Results highlight some differences in both tests. The presented analysis is informative for both ontology curators and users of functional enrichment methods.

%and GOA ..... and formulation of IC provided by. We calculated IC of all GO Terms for each GO version yielding to xxx calclation. We an in-depth study of the underlying changes and their impact on the analysis results. Results demonstrated that....

%\textbf{Paper is structured as follows.}

\section{\bf Materials and Methods}
\label{sec:material}

\subsection{IC Calculation}

This section discusses the state of the art approaches to compute the information content of a given term belonging to an ontology, presenting some examples related to the biological context. We aim to present the theoretical basis of this calculation in order to highlight the possible causes of bias that we present in next sections.

Term information content (IC) approaches can be divided into two families as we depict in Figure \ref{fig:intrisicvsestrinsic}: extrinsic ( or annotation based) and intrinsic  (or topology-based) IC approaches. 

While intrinsic approaches exploit only the intrinsic topology of the GO graph thus they are only subject to the variation of GO structure, the annotation-based approach requires the addition of annotation data for the corpus under consideration thus they present more causes of biases due to the variability of both GO structure and corpora of annotation.

Intrinsic IC calculus can be estimated using different topological characteristics as ancestors, number of children, depth (see \cite{harispe2013framework} for a complete review).  For instance, the classical formulation provided in the seminal paper of Resnick\\ \cite{resnink:simmeasure:879855} calculates the IC of a concept by considering all the top-down path from a concept $a$ to the reachable leaves, namely $p(a)$, and then calculate the log of this number yielding to the formula: \begin{equation}
 -log(p(a)).
 \end{equation}

Obviously, the growth of the reachable path determines the increase of the IC. One of the problems of this simple formulation is that it takes value in the interval $0, \infty $.

Instead, more recently has been proposed a normalized variant in which the maximum IC for all concept is used as normalizing factor as follows:\begin{equation}
IC_{Resnik}(a) =-log\left(\frac{p(a)}{max\_pa})\right)
\end{equation}
where \emph{max\_pa} is max p(a) for all concepts.

Seco et al. \cite{14755292} compute the IC of a concept as the ratio between the number of hyponyms in ontology (i.e. the number of descendant) with respect to the whole  number of ontological concepts yielding to the following equation.
\begin{equation}
IC_{Seco\,et\,al.}(a) =\frac{log\left( \frac{hypo(a)+1}{max\_nodes} \right)}{log\left( \frac{1}{max\_nodes} \right)}
\end{equation}

With respect to the Resnik formulas, according to Seco et al. concepts with many hyponyms are less informative than leaves of DAG, thus, if two concepts at different level of generality in DAG have a equal number of hyponyms, they are considered equally informative.

One of the drawback of this formulation is that the relative position of a concept with respect to the maximum depth of the taxonomy is not considered. Thus Zhou et al. \cite{zhou2008new} add to Seco's approach the depth of concepts in the taxonomy \emph{depth(a)} and the maximum depth of the taxonomy  \emph{max\_depth}.
\begin{equation}
IC_{Zhou\,et\,al.}(a) =k-\left(1- \frac{log(hypo(a)+1)}{log(max\_nodes)} \right)+(1-k)\left(\frac{log(depth(a))}{log(depth\_nodes)} \right)
\end{equation}
where K is factor which enables to weigh the contribution of the two evaluated features.

The IC of terms as proposed in Sanchez et al. \cite{sanchez2011ontology} exploits only the number of leaves and the set of  ancestors of a including itself, \emph{subsumers(a)} and introduce the root node as number of leaves \emph{max\_leaves} in IC assessment. Leaves are more informative than concepts with many leaves, roots, so the leaves are suited  to describe and to distinguish any concept.
\begin{equation}
IC_{Sanchez\,et\,al.}(a) =-log\left(\frac{\frac{|leaves(a)|}{|subsumers(a)|}+1)}{max\_leaves+1} \right)
\end{equation}

In order to achieve a normalized measure, this formula may be adapted  normalize considering spirit formulated in Faria. et al as proposed in \cite{Faria2007} yielding to: \begin{equation}
IC_{SanchezAdapted\,et\,al.}(a) =-log\left(\frac{|leaves(a)|+1)}{max\_leaves+1} \right)
\end{equation}

Harispe et al.  revise  the IC assessment  suggested by Sanchez et al. considering \emph{leaves(a)} = \emph{a} concept when \emph{a} is a root and evaluating \emph{max\_leaves} as the number of inclusive ancestors of a node. In this way, the specificity of leaves according to their number of ancestors is distinguished.
\begin{equation}
IC_{Harispe\,et\,al.}(a) =-log\left(\frac{\frac{|leaves(a)|}{|subsumers(a)|})}{max\_leaves} \right)
\end{equation}

\subsection{IC-Based Semantic Similarity Measures}

Semantic similarity is a function to measure closeness among terms belonging to the same ontology \cite{Guzzi2012}. There exist different classification of semantic similarity, for instance Guzzi et al. proposed in a recent work to classify measures according to whether or not
they consider some aspects or use some common
strategies in : (i) Term Information Content (IC), (ii) Term Depth, (iii) based on a common ancestor,
(iv) based on all common ancestors, (v) Path
Length and (vi) Vector Space Models (VSM).

Considering the analysis of proteins, it has been demonstrated  that best performances in terms of assessment with respect to biological features are obtained by IC-based measures \cite{Guzzi2012,Pesquita2009,cho2013m}.

 Resnik's method \cite{Resnik1995} computes the semantic similarity between $t_1$ and $t_2$ by the greatest information content of common ancestor terms of $t_1$ and $t_2$. In other words, this method estimates the specificity of the most specific common ancestor term (SCA).
\begin{equation}
    sim_{\mbox{\tiny \emph{Resnik}}}(t_1, t_2) = \max_{t_0 \in C(t_1, t_2)} (-\log P(t_0)),
\end{equation}
where $C(t_1, t_2)$ is a set of all common ancestor terms of $t_1$ and $t_2$. 

Lin's method \cite{citeulike:1238} normalizes Resnik's method by the average information content of $t_1$ and $t_2$.
\begin{equation}
    sim_{\mbox{\tiny \emph{Lin}}}(t_1, t_2) = \max_{t_0 \in C(t_1 , t_2)} \bigg(\frac{2\times \log P(t_0)}{\log P(t_1)+\log P(t_2)}\bigg).
\end{equation}

Jiang's method \cite{Jiang1997Semantic} computes the sum of differences of the information contents between SCA and the input GO terms, $t_1$ and $t_2$ applying an approach similar to Jaccard's Index.
\begin{equation}
    sim_{\mbox{\tiny \emph{Jiang}}}(t_1, t_2) = \frac{1}{\min_{t_0 \in C(t_1 , t_2)} (2\times \log P(t_0)-\log P(t_1)-\log P(t_2)) + 1}.
\end{equation}

Schlicker et al. \cite{Schlicker01012010} proposed a combined method of Resnik's and Lin's methods, which is called simRel. If SCA is defined as the term where two paths towards the root from $t_1$ and $t_2$ converge, multiple SCAs of $t_1$ and $t_2$ generally occur in a DAG structure since each GO term has multiple parent terms. Couto et al. \cite{Couto2007} defined a set of all SCAs of pairwise paths towards the root from $t_1$ and $t_2$ as common disjunctive ancestors. They proposed add-on semantic similarity methods, GraSM which averages the information contents of common disjunctive ancestor terms and DiShln which is a slight modification of GraSM \cite{Couto:SEMANTICSIMILARITIBIOMEDONTO:PLOS2009}.

Finally, many integrative approaches of two different categories have recently been proposed to achieve higher accuracy in measuring functional similarity of proteins. For example, Wang et al. \cite{citeulike:7730206} proposed a combination of the normalized common-term-based method and the path-length-based method. Their semantic similarity measure, called G-SESAME, scores a protein pair by the common GO terms having the annotations of the proteins, but gives different weights to the common GO terms according to their depth. Pesquita et al. \cite{Pesquita2008} proposed simGIC which integrates the normalized common-term-based method with information contents. Instead of counting the common terms, simGIC sums the information contents of the common terms.
\begin{displaymath}\label{eq:simgic}
    sim_{\mbox{\tiny \emph{simGIC}}}(t_1, t_2) = \frac{\sum_{t_i \in C(t_1)\cap C(t_2)} \log P(t_i)}{\sum_{t_j \in C(t_1)\cup C(t_2)} \log P(t_j)},
\end{displaymath}
where $C(t_1)$ is a set of all ancestor terms of $t_1$.

Finally, two recent IC based measures were proposed by Cho et al. \cite{cho2013m}. The rationale is to  integrate two orthogonal features. Since Resnik's method computes the information content of SCA of two GO terms $t_1$ and $t_2$, it focuses on their commonality, not a difference between them. In contrast, Lin's and Jiang's methods measure their difference only.  %In particular, Lin's model reflects a significant bias towards higher scores when the set of annotating proteins on SCA is similar to those on $t_1$ and $t_2$. This case commonly occurs in GO because of the shallow annotation problem \cite{Guzzi2012}. To enhance the performance of these annotation-based methods, we normalize Resnik's semantic similarity of $t_1$ and $t_2$ by their distance.

simICNP (Information Content of SCA Normalized by Path-length of two terms) uses the information content of common ancestors normalized by the shortest path length between $t_1$ and $t_2$ as the distance.
\begin{equation}\label{eq:icnp}
    sim_{\mbox{\tiny \emph{ICNP}}}(t_1, t_2) = \frac{-\log P(t_0)}{len(t_1,t_2) + 1},
\end{equation}
where $t_0$ is SCA of $t_1$ and $t_2$. This method gives a penalty to Resnik's semantic similarity if $t_1$ and $t_2$ are located farther from their SCA. 

simICND (Information Content of SCA Normalized by Difference of two terms' information contents) employs the information content of SCA normalized by the difference of information contents from the two terms to SCA, as Jiang's method uses.
\begin{equation}\label{eq:icnd}
    sim_{\mbox{\tiny \emph{ICND}}}(t_1, t_2) = \frac{-\log P(t_0)}{2\cdot \log P(t_0) - \log P(t_1) - \log P(t_2) + 1}.
\end{equation}
This method gives a penalty to Resnik's semantic similarity if the information contents of $t_1$ and $t_2$ are higher than the information content of their SCA. 

%
%
%

%   \begin{table*}
%   \centering
%  \caption{}\label{tab:ieastat}			
%   \begin{tabular}{|p{1.5cm}|p{1.5cm}|p{1.5cm}|p{1.5cm}|p{1.5cm}||p{1.5cm}||p{1.5cm}|p{1.5cm}|}
%			 \hline
%\tiny MEASURE  &  \ \tiny AVERAGE & \ \tiny SD  &  \ \tiny MEDIAN & \ \tiny FIRST \tiny QUARTILE &  \ \tiny SECOND \tiny QUARTILE & \ \tiny THIRD \tiny QUARTILE & \ \tiny P-VALUE \\ \hline
%
%\tiny Harispe & \tiny 12,135 &	\tiny 1,212	 & \tiny 12,277 & \tiny 11,871	 & \tiny 12,277 &	\tiny 12,788 &	\tiny $p-value < 2,2e-16$ \\ \hline
%\tiny Resnick  & \tiny 0,999	& \tiny 0,005 &	\tiny 1 &	\tiny 1 &	\tiny 1 &	\tiny 1 & \tiny $p-value<2,2e-16$ \\ \hline
%\tiny ResnickUnpro & \tiny 0,935 &	\tiny 0,110 &	\tiny 1 &	\tiny 0,896 &	\tiny 1 &	\tiny 1 & \tiny $p-value<2,2e-16$ \\ \hline
%\tiny SANCHEZ  & \tiny 9,908 &	\tiny 0,518 & 	\tiny 10,08 &	\tiny 9,962  & 	\tiny 10,08	& \tiny 10,08 & \tiny $p-value<2,2e-16$ \\ \hline
%\tiny SanchezAdapted  & \tiny 0,896 & \tiny 0,080 & \tiny 	0,931 &	\tiny 0,891 &	\tiny 0,931 &	\tiny 0,931 & \tiny $p-value<2,2e-16$\\ \hline
%\tiny Seco  & \tiny 0,935 &	\tiny 0,110 &	\tiny 1 &	\tiny 0,896 &	\tiny 1 &	\tiny 1  & \tiny $p-value<2,2e-16$\\ \hline
%\tiny Zhou & \tiny 0,833 & \tiny 0,077  & \tiny 0,836 &	\tiny 0,807 &	\tiny 0,836 &	\tiny 0,880 & \tiny $p-value<2,2e-16$\\ \hline
%
%\end{tabular}
%\end{table*}
%

%\end{document} 

\subsection{Studying GO Evolution }
\subsubsection{Definition of GO Evolution}

Ontology changes may be distinguished on two main classes: (i) changes on the structure of GO, (ii) changes on annotation corpora.

Regarding the first class, it should be noted that different studies have provided different way of classification of changes. We here follow the classification proposed in Pesquita and Couto \cite{pesquitaPredictionofChanges} based on previous work of Flouris et al. \cite{flouris2008ontology}. In that work ontology evolution is defined as the  process of modifying an ontology in response to a certain change in the domain or its conceptualization. Changes are related to: (i) modification of the real word modeled by ontologies (e.g. novel experiments that demonstrates novel relation among biological concepts), (ii) a reconsideration of is\_a relations among elements of  the ontology, (iii) the extension of the scope of the ontology by adding novel information previously unavailable, and (iv) the correction of previous mistakes on the structure and on annotations. It should be noted that differently from the other fields, the high dynamic of biological field determines that the majority of changes on the ontology are within the third and fourth class.

For instance, considering the version of GO of 2011-06-11 and 2011-06-18 we may report following changes (the complete list is available at - \url{http://www.gene-ontology.org/internal-reports/ontology/2011-06-18/weekly-2011-06-18.txt} -).  Tables \ref{tab:examplesofadd} and \ref{tab:examplesofstructuremodification} report respectively a summary of novel terms, deleted terms and of changes in GO Structure.

\begin{table}

\caption{Examples of GO Changes. Insertion of novel GO Terms in GO version of 2011-06-18 }
\label{tab:examplesofadd}
\centering
\begin{tabular}{|p{11cm}|}
\hline 
Insertion of Novel GO Terms \\ \hline
GO:0002185	creatine kinase complex	cellular\_component \\
GO:0002186	cytosolic creatine kinase complex	cellular\_component \\
GO:0002187	mitochondrial creatine kinase complex	cellular\_component \\
GO:0035888	isoguanine deaminase activity	molecular\_function \\
GO:0035889	otolith tethering	biological\_process \\
GO:0035890	exit from host	biological\_process \\
GO:0035891	exit from host cell	biological\_process \\
\hline
\end{tabular}
\end{table}

\begin{table}
\caption{Examples of GO Changes. $-$ and $+$ mean respectively the elimination\ insertion of an is\_a relationship.}
\label{tab:examplesofstructuremodification}

\centering
\begin{tabular}{|p{3cm}|p{9cm}|}
\hline 
Type & GO Terms Involved \\ \hline
Modification on structure & GO:0009962 : regulation of flavonoid biosynthetic process
\- is\_a: GO:0043455
+ is\_a: GO:2000762 \\ \hline
Modification on Structure & 
GO:0031537 : regulation of anthocyanin metabolic process
\- is\_a: GO:0031323
\- is\_a: GO:0043455
+ is\_a: GO:2000762
\\ \hline
\end{tabular}

\end{table}

\begin{figure}
\centering
\subfigure[Evolution of CC Ontology]
   {\includegraphics[width=1.2\textwidth]{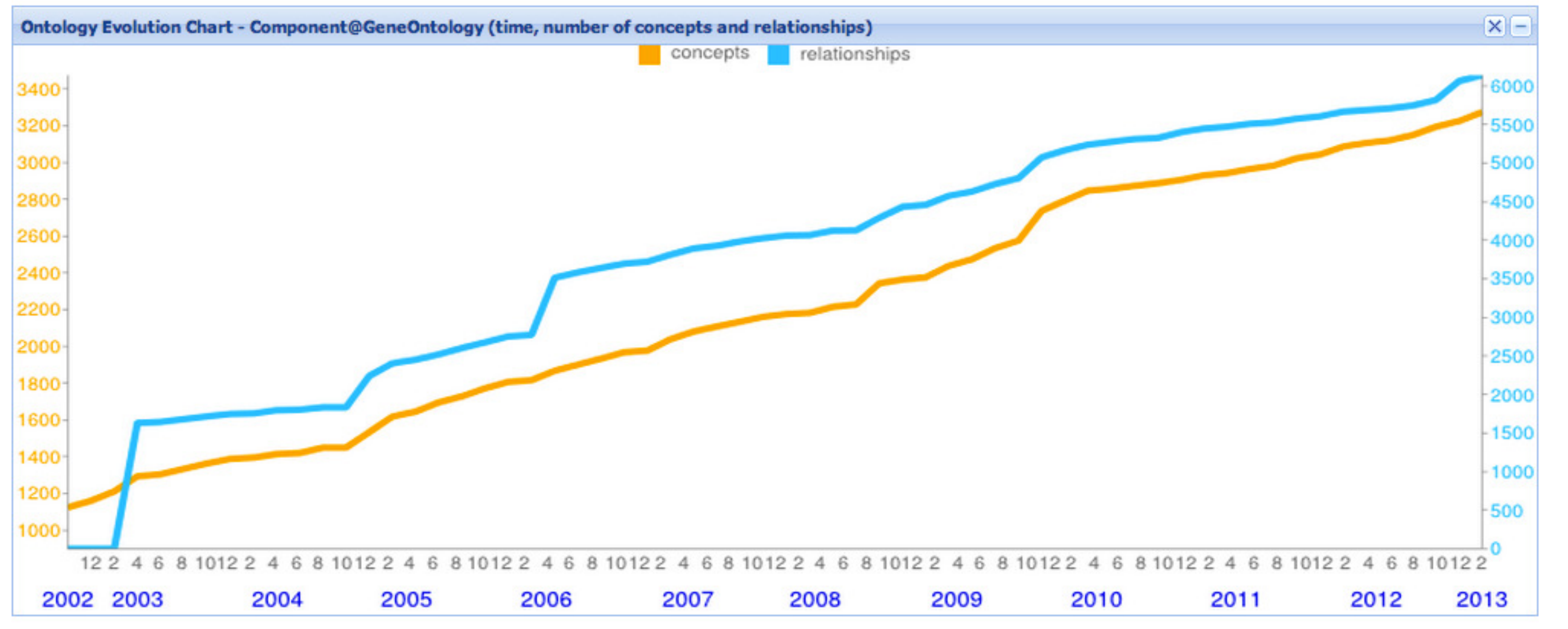}}
 \hspace{5mm}
 \subfigure[Evolution of BP Ontology]
   {\includegraphics[width=1.2\textwidth]{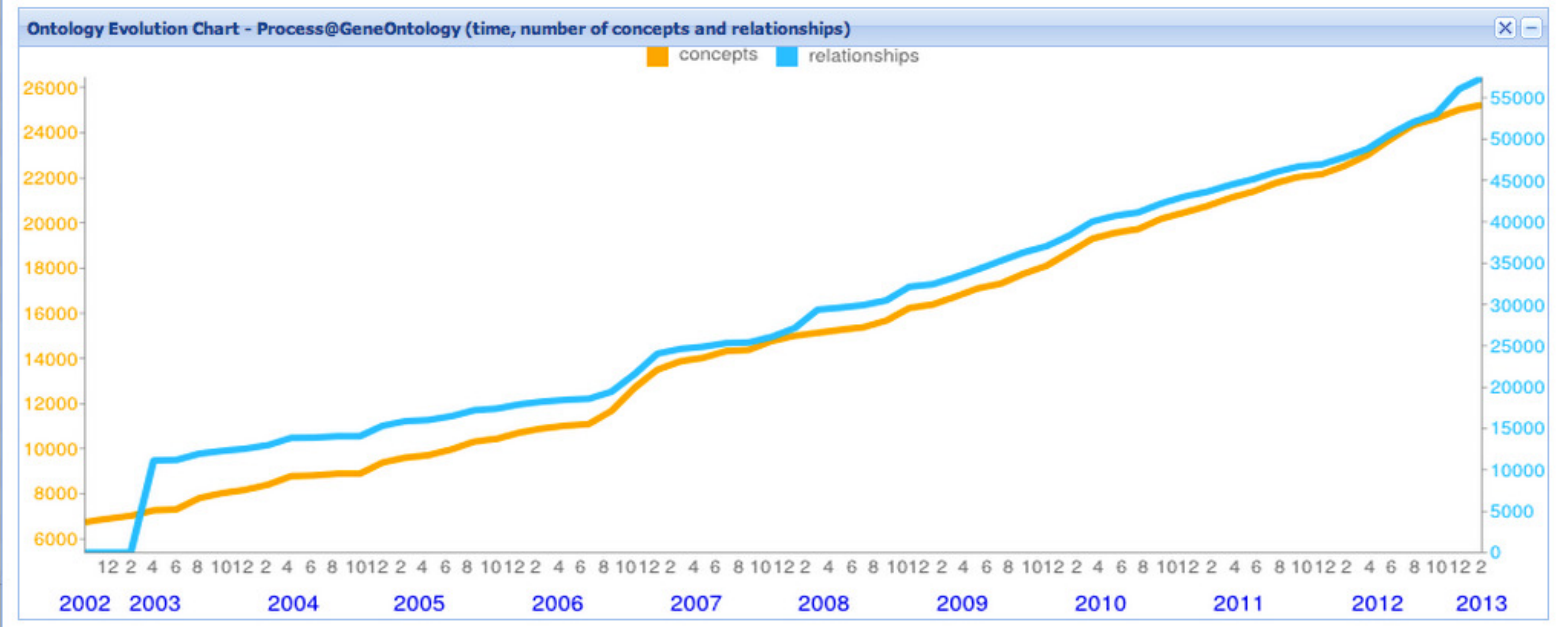}}
\subfigure[Evolution of MF Ontology]
   {\includegraphics[width=1.2\textwidth]{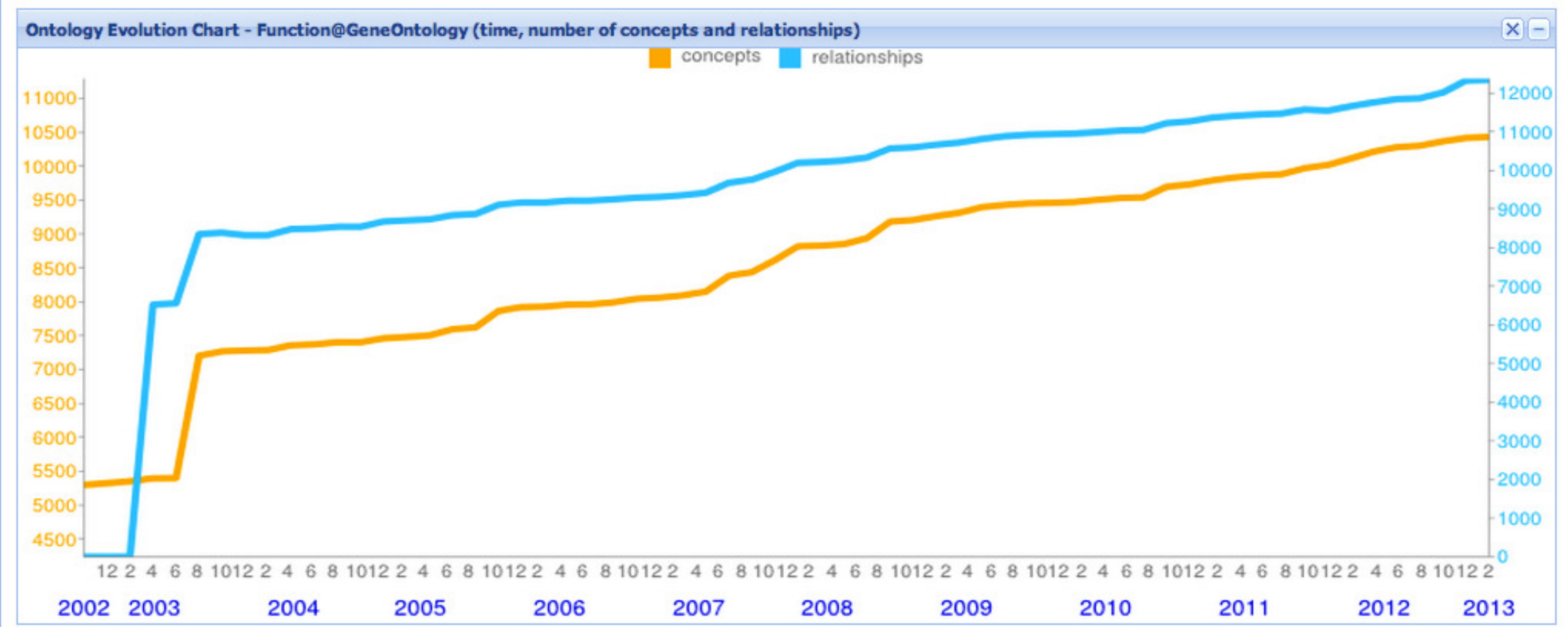}}
\caption{Evolution of taxonomies in GO as reported by Onex. }
\label{fig:evolution}
\end{figure}

The Gene Ontology Consortium provides periodically a summary of changes in terms of added-deleted and modified terms that is available on the web at \url{http://www.geneontology.org/internal-reports/ontology/}. Figure \ref{fig:evolution} summarizes these changes from 2005 to 2012.

%\begin{table}
%\label{tab:goevostats}
%\caption{Summary of Changes in GO as reported by GO Consortium.}
%
%\textbf{QUI INSERISCO QUALCHE NUMERO COME TABELLA PER DARE L'IDEA DEI CAMBIAMENTI. anche solo ad esempio dal 2006 al 2011 cercando di mostarre}
%prendere come dato di riferimento quello contenuto in:
%http://dbserv2.informatik.uni-leipzig.de:8080/onex/
%\end{table}

Independently from the causes, the evolution of the ontology comprises three basic operations: add, remove or modify. Considering the add operation, we may evidence three evolutions of the ontology: ontology extension, ontology refinement and ontology enrichment. Ontology extension, as reported by Pesquita and Couto\\ \cite{pesquitaPredictionofChanges}, is defined as    the process by which new single elements are added to an existing ontology. Ontology extension regards the changes due to the addition of novel elements motivated, for instance, by novel discoveries. 

Ontology refinement is the addition of new concepts to an ontology, and the subsequent adding of subsumption relations on the ontology.

Ontology enrichment regards the adding of non-taxonomical relations (i.g. GO \textit{regulates} ) or other axioms. For instance as reported in \cite{pesquitaPredictionofChanges} the addition of the relation âregulates" between the GO concepts âregulation of mitochondrial translation" and âmitochondrial translation".

The whole set of changes determines thus a remarkable modification in the whole GO.  Hartung et al.\cite{HartungOndex} provided a more formal and compact tool, namely OnEx (Ontology Evolution Explorer), that it is able to determine the  \textit{semantic diff} of changes among versions. The tool is  able to highlight visualization of changes but they do not provide interpretation of changes nor the impact. Here we report the trend of the evolution of GO as reported in the OnEx web site available at \url{http://dbserv2.informatik.uni-leipzig.de:8080/onex/}. Figure \ref{fig:evolution} summarizes changes in CC, BP, and MF ontology (all the images are extracted from the web site:\url{http://dbserv2.informatik.uni-leipzig.de:8080/onex/}).

Each GO Terms  may be associated with any number of gene products. These associations are known as 'annotationsâ and can be created either manually or automatically \cite{duPlessis}. Annotation may be made by a curator on the basis of the analysis of literature providing detailed and specific information. Automatic (or electronIcally infErred Annotation -IEA), are made using algorithms that consider gene product properties, such as orthology, domains and sequence similarity \cite{Guzzi2012}. They in general provide a broad coverage of annotation and cover a significantly larger field of knowledge.
GOC provides over 200 million annotations stored in the Gene Ontology Annotation Database \cite{GOA:citeulike:461337} with around 99\% of these being automatically created. The annotation database is periodically updated and the trend, evidenced in Table \ref{tab:GOAtrend} demonstrates a constant increase in the number of annotations.

 \begin{table}

 \centering
 \caption{Trend of Evolution of GO Annotations considering Uniprot database as reported in Pesquita and Couto Plos Comp Bio 2012.}
 \label{tab:GOAtrend}
 \begin{tabular}{|p{3cm}|c|c|}
 \hline 
 GO &	TOTAL 	&	MANUAL 	\\ 
 VERSION & ANNOTATIONS & ANNOTATIONS \\
 \hline
Jan 2005	&	6.0 M	&	0.50 M	\\ \hline
Jul 2005	&	7.1 M	&	0.62 M	\\ \hline
Jan 2006	&	7.3 M	&	0.56 M	\\ \hline
Jul 2006	&	9.0 M	&	0.56 M	\\ \hline
Jan 2007	&	10.4 M	&	0.62 M	\\ \hline
Jun 2007	&	12.4 M	&	0.66 M	\\ \hline
Jan 2008	&	19.0 M	&	0.73 M	\\ \hline
Jul 2008	&	23.0 M	&	0.78 M	\\ \hline
Jan 2009	&	24.7 M	&	0.79 M	\\ \hline
Aug 2009	&	33.0 M	&	0.87 M	\\ \hline
Jan 2010	&	33.5 M	&	0.91 M	\\ \hline
Jul 2010	&	60.5 M	&	1.06 M	\\ \hline
Jan 2011	&	54.4 M	&	1.23 M	\\ \hline
Jul 2011	&	63.8 M	&	1.35 M	\\ \hline
Jul 2012	&	77.8 M	&	1.41 M	\\ \hline
 
  \hline 
 \end{tabular}

 \end{table}

\begin{figure}[tb]
	\begin{centering}
	\includegraphics[width=1.2\textwidth]{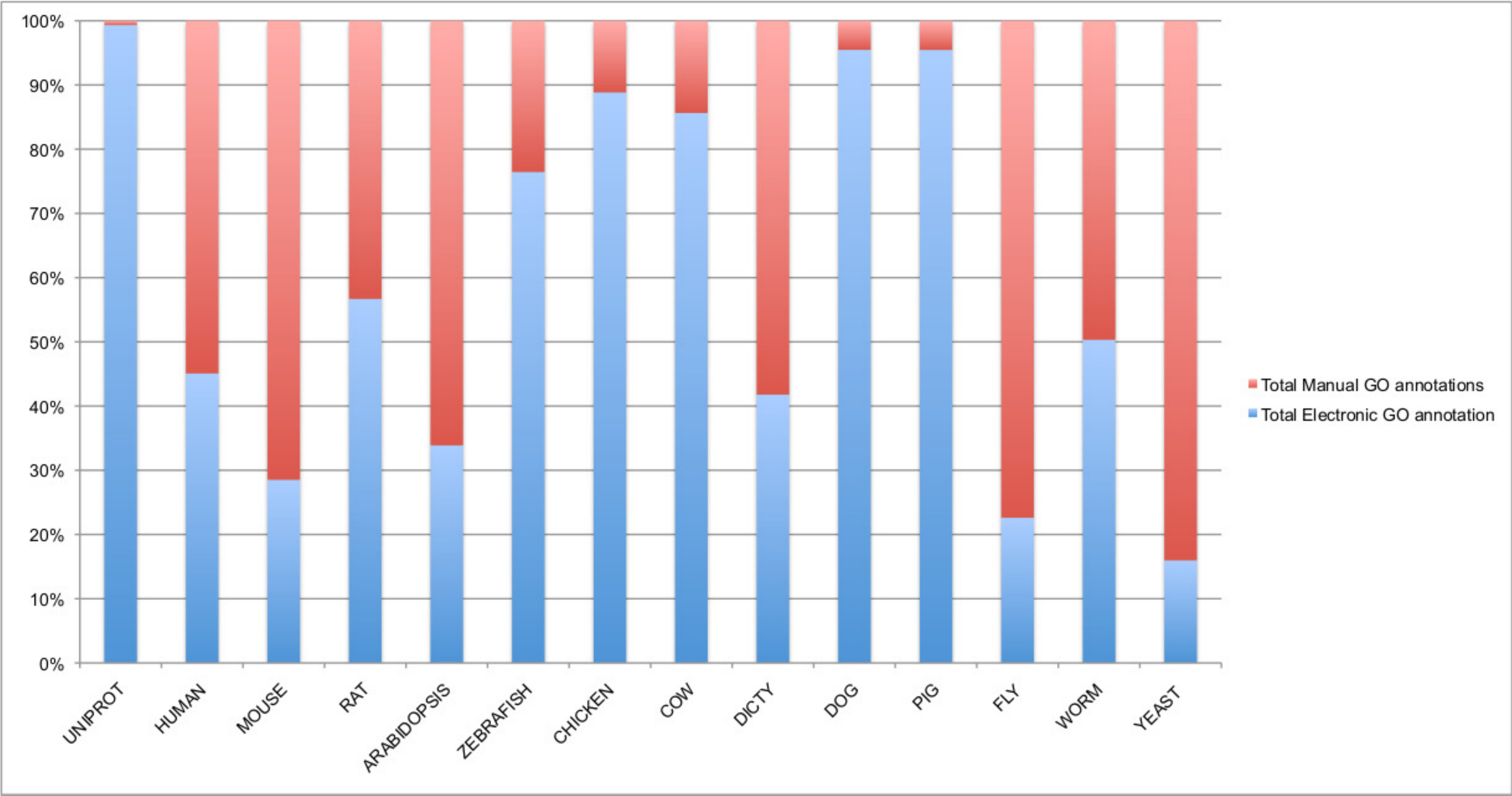}
	\caption{Ratio of IEA vs non-IEA Annotations on Different Species}
	\label{fig:ieastats}
	\end{centering}
\end{figure}

Manuals annotations are in general more precise \cite{Guzzi2012} and specific than IEA ones. Unfortunately their number is in general lower (as shown in Figure \ref{fig:ieastats} ) and this ratio is variable.

A considerable number of genes and proteins is annotated with generic GO terms (this is particular evident when considering novel or not well studied genes). The role of these general annotations is to  suggest the area in which the proteins or genes operate. This phenomenon  affects particularly IEA annotations derived, from instance, from literature.

Similarly to changes in GO structure, the annotation corpora are in continuos evolution to reflect ongoing work, novel discoveries or the introduction of novel algorithm that may discover novel annotation or demonstrate the inconsistency of the existing ones.

%Many changes are made to both the ontology and annotation sets over time - some of these changes are planned and announced by GOC or its members via mailing lists or release notes [10,12], whereas others are not and reflect ongoing improvements, such as user requests for updates to the ontology [13] or annotations [14], as well as revisions in response to quality assurance checks.

%and are critical for creating learning sets for automated pipelines. Automatic annotations are made using algorithms that consider gene product properties, such as orthology, domains and sequence similarity [5,9], and they provide a broad coverage of annotation and cover a significantly larger taxonomic range than manual annotations. This difference in coverage is illustrated by the annotation statistics from the database of the UniProt GO Annotation project (UniProt-GOA) that includes GO annotation from all of the GOC members [5]; as of November 2013, GOC provides over 200 million annotations, with around 99% of these being automatically created [10,11].

\subsubsection{Studies on GO Evolution}

The study of the evolution of the GO has been performed in the past by many authors yielding to the introduction of both formal theories to: (i) describe ontology changes, (ii) to measure the quality of ontology (e.g. how the ontology mimics the reality), (iv) to evaluate the impact on the quality of the ontology, (v) to develop tools for such studies.

The paper by Leonelli et al. 
\cite{Leonelli:GOEVOLUTION} provides a formal study on the motivation of the changes in the ontology. They identified five major causes of changes:  (1) the discovery  of anomalies within GO (i.e. the misuse of a term); (2) the broadening of the coverage of the scope of GO; (3) the presence of a different use of the same GO Term  across multiple user communities; (4) the presence of novel discoveries that cause the change of the meaning of a term as well as of the relations among terms; and (5) the broadening of the range of non-taxonomical relations. The paper focuses mainly on the determination of a formal framework to improve the corrispondence among GO terms and biological knowledge without analysing the impact of changes.

%%%%%%%% QUALITY OF CHANGES

Ceusters \cite{ceusters2009applying} analyzed changes between 2001 and 2007 for measuring to what extent the structure of a terminology mimics reality. Author reports that  the quality of the BP, CC and MF branches of the GO increased, and best results were achieved in MF. He also observed that the increase of the size of GO in terms of number of GO terms is in general correlated with an increase of quality. Results are in contrast with those reported by Dameron et al. \cite{complexityGOEvolution} that showed that the complexity increased for BP, decreased slightly for CC and remained stable for MF.

Alterovitz et al. \cite{alterovitz2010ontology} looked at the distribution of information content on GO terms and they propose an \textit{ontology engineering} methodology, i.e an information theory-based approach to automatically organize the structure of GO and optimize the distribution of the information within it. The method is sound in principle but it has not been applied in the practical evolution, therefore our analysis remains still valid.

KÃ¶hler et al. defined a formal method to define and analyze the quality of the definition of GO Terms \cite{kohler2006quality}. Mungall et al proposed a method to improve quality of annotation by detecting missing annotation as well as incorrect ones by using description logic and automatic reasoning \cite{mungall2011cross}. 

Faria et al proposed a way to improve annotation consistency by using association rules, but, to the best of our knowledge, they do no consider information content since they manually remove low informative terms \cite{faria2012}.

%%%%%%% TENTATIVI PER MIGLIORARE

Existing works that studies to what extent modifications of the GO and of gene annotations databases impacted on subsequent analysis main focused on gene enrichment analysis. Gross et al. \cite{Gross15102012} studied the impact of changes on classical gene enrichment algorithms, i.e. the description of  experimental data by sets of GO terms. Main results of this work are: (i) the deminstration that ontology changes are unequally distributed among the structure ant that they may be clustered into regions representing specific topics, (ii) these changes do not always modify the result of term enrichment analyses since the terms are often semantically related. Dameron et al. \cite{complexityGOEvolution} considered there results and demonstrated that for BP, most modifications occurred deep into the hierarchy. Therefore it is also possible that term enrichment analyses return sets of more general GO terms that are more stable.

Clark et al. \cite{loguercio:TASKBASED} proposed a model to evaluate the quality and the completeness of GO annotations by applying a task-based approach. In particular they focused on different task belonging to  gene enrichment analysis class. They focused on the quality of annotations, without considering changes in the structure, whereas we focused on GO proper. It should be noted that intrinsic information content may be affected only by structural changes while extrinsinc information content are sensible to changes on annotation corpora.

Pesquita and Couto \cite{pesquitaPredictionofChanges} proposed a semi-automatic approach for monitoring changes and for predicting possible needed changes. They applied it to  GO over the 2005â2010 period to predict the portions of GO that would be extended.  The focus of this study was the analysis of novel classes and its relation with respect to existing ones. By the analysis of classes depth, and their ancestor and childres they determined if new classes provide a finer description or cover a new domain. One of the conclusion of this work is that  in BP, CC and MF, the majority of new subclasses are added as children of non-leaf classes (therefore this change has a great impact on information content of existing terms).  They also observed that the refinement of CC and MF occurs mostly via single insertions, whereas in BP, groups of related classes are inserted together.

%%%%%%% TOOLS

Park et al. \cite{park2008monitoring} developed a set of visualization methods based on a layered and colored graph to highlight changes among two version of GO.  Hartung et al.\cite{HartungOndex} provided a more formal and compact tool, namely OnEx (Ontology Evolution Explorer), that it is able to determine the  \textit{semantic diff} of changes among versions. Both tools are able to highlight visualization of changes but they do not provide interpretation of changes nor the impact.

%\subsection{Semantic Similarites and Information Content}

%\subsubsection{IC-based Semantic Similarities}
%\input{semsim}

%\textbf{Descrivere l'esperimento}

\section{\bf Results}
\label{sec:results}
%\subsection{Ontology Changes as revealed by Codex}
%\input{codex}

%\textbf{Qui mettiamo tutti i risultati corredando le parole con ampie tabelle e grafici.}

\subsection{IC Changes}

We here analyzed in detail the distribution of different ICs. Table \ref{tab:experiments} summarizes main parameters of the experiment reporting the GO version and IC we used. As introduced before for each GO version we calculated the IC of all the GO terms using all the cited formulations of IC.
\begin{table}
\label{tab:experiments}
\caption{Parameters of the Experiment}
\centering
\begin{tabular}{|p{2.5cm}|p{3cm}|}
\hline 
GO Version\footnote{Release of April} & IC Measures \\ \hline
2006	2007	2008	2009	2010	2011	2012	2013	2014 & Zhou,
Sanchez Adapted,
Sanchez,
Harispe.
\\ \hline
\end{tabular}
\end{table}
For each GO we computed the IC of all the contained terms by applying all the measures (data are available on the web site of the project \url{https://sites.google.com/site/evolutionofic/} ). As preliminary test we verified that no one  distribution follows a Gaussian model by applying a Pearson's chi-square test for normality Test (p-value less than 0,05 for each distribution). On the basis of this consideration we decided to use non-parametric test for following comparison.

As initial step we performed an analysis within the same measure, in order to evidence that distributions of ICs of different years are different. For these aims we used the Wilcoxon test that is a nonparametric test designed to evaluate the difference between two treatments or conditions where the samples are correlated. In particular, it is suitable for evaluating the data from a repeated-measures design in a situation where the prerequisites for a dependent samples t-test are not met.

\paragraph{Comparison within Measure}

Results confirmed that there exist a significant difference within the same measure considering different years (results are not reported here for clarity - see supplemental materials). Differences among years are evidenced in Figure \ref{fig:measurebyyear}.

\begin{figure}[ht]
\centering
%\includegraphics[width=4in]{ intrinsic.pdf}
%\subfigure[Zhou IC]
   {\includegraphics[width=1\textwidth]{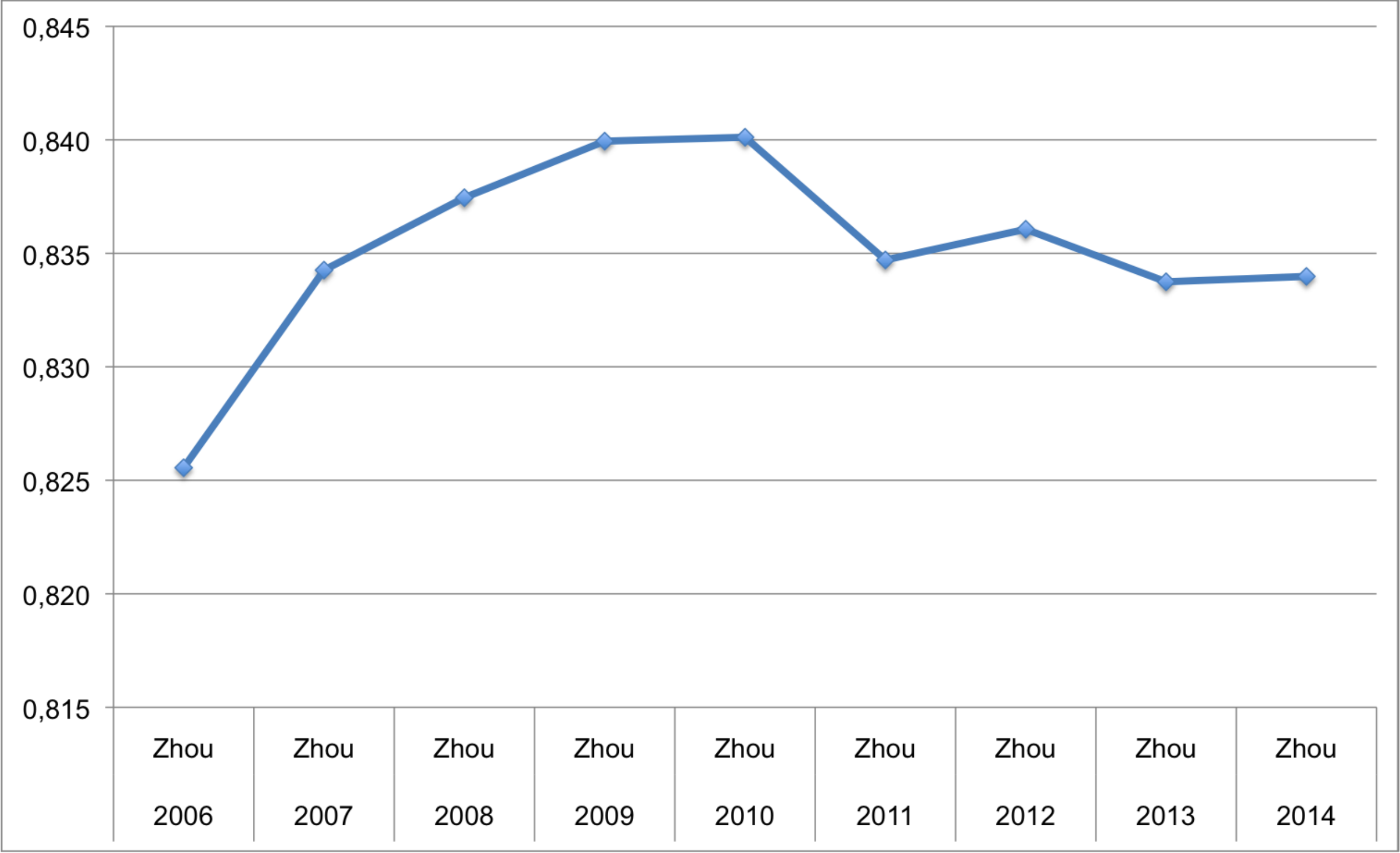}}
   \caption{Evolution of median IC values on GO.}
\label{fig:measurebyyearszhou}
\end{figure}
 %\hspace{4em}
\begin{figure}[ht]
%\subfigure[Sanchez IC]

   {\includegraphics[width=1\textwidth]{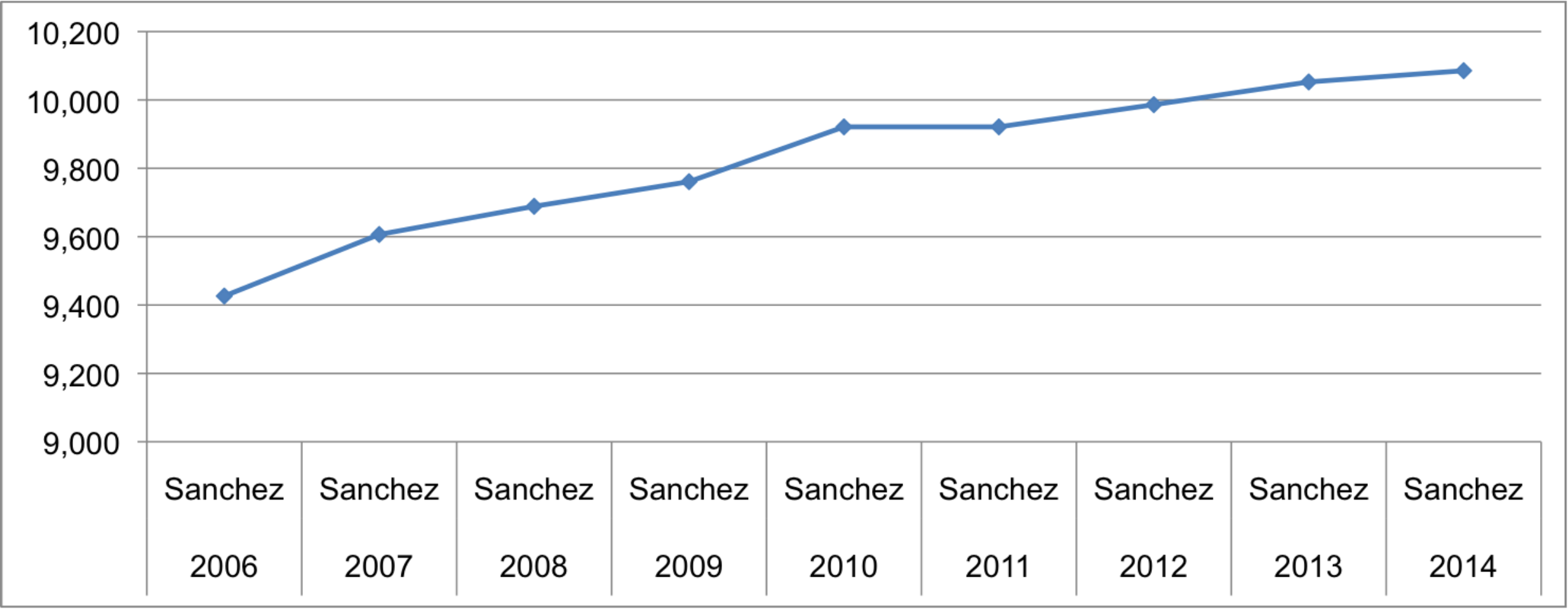}}
 %\hspace{4em}
 \caption{Evolution of median IC values on GO.}
\label{fig:measurebyyearsan}
\end{figure}
%\subfigure[Sanchez Adapted IC]
\begin{figure}[ht]
   {\includegraphics[width=1\textwidth]{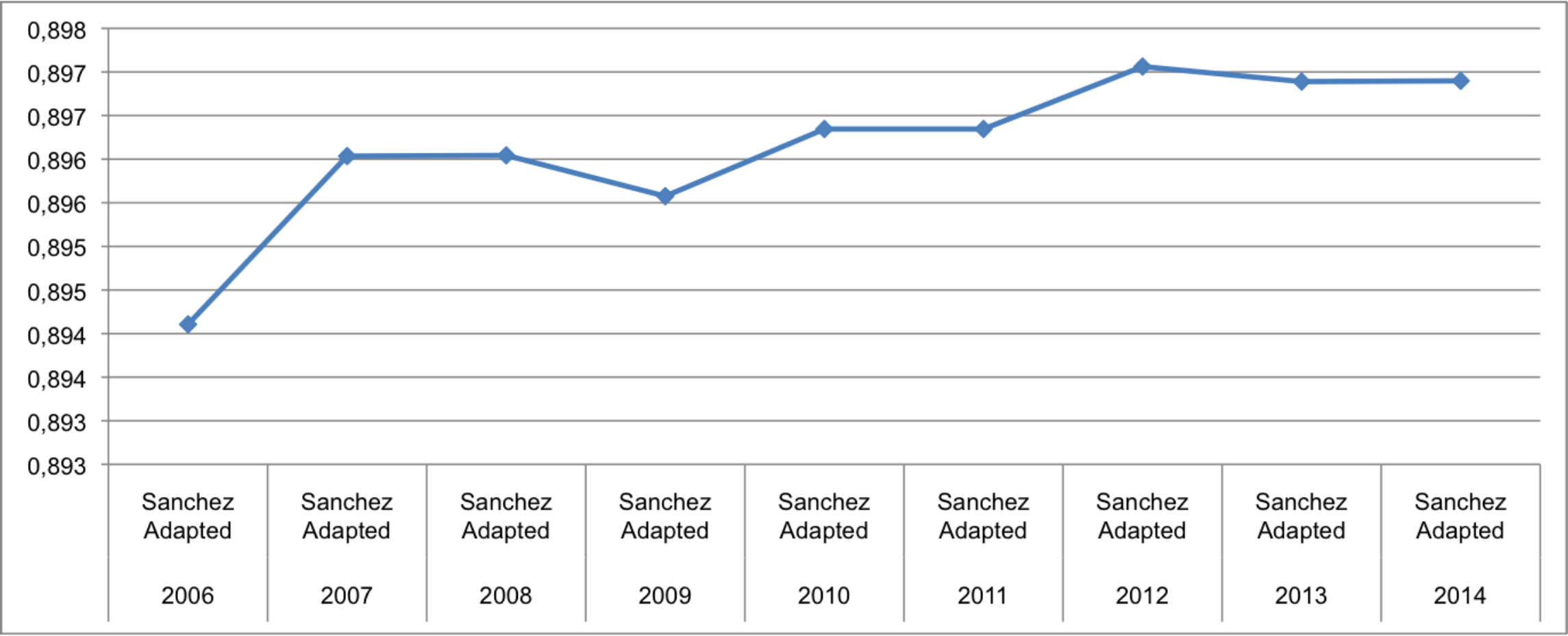}}
\caption{Evolution of median IC values on GO.}
\label{fig:measurebyyearsanad}
   \end{figure}
% \hspace{4em}
%\subfigure[Harispe IC]
\begin{figure}[ht]
   {\includegraphics[width=0.8\textwidth]{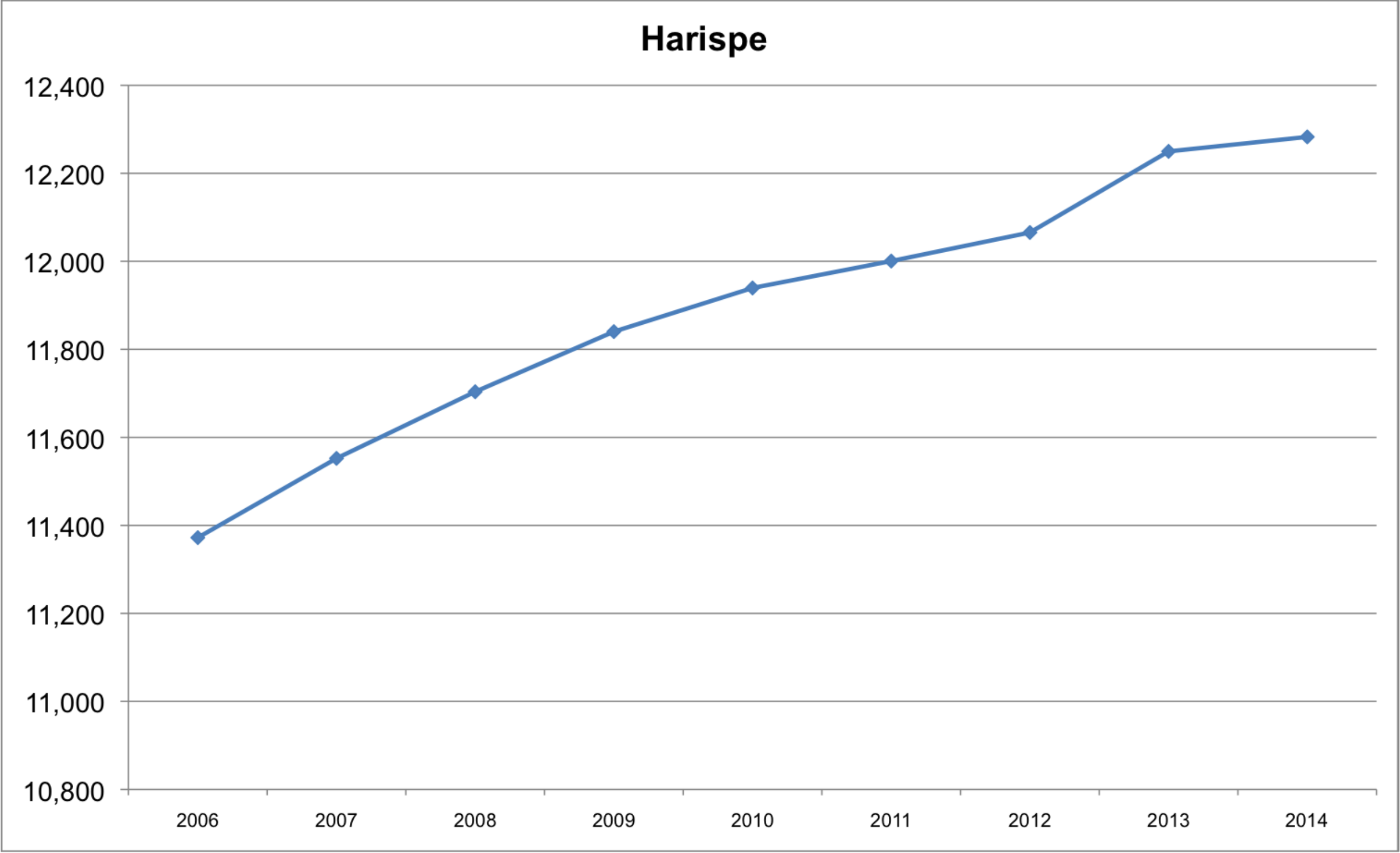}}
\caption{Evolution of median IC values on GO.}
\label{fig:measurebyyearharispe}
\end{figure}

\noindent\paragraph{\textbf{Comparison within year}}

Moreover we performed a comparison among the same year by comparing different IC formulation. In order to simplify the number of comparison we selected one GO for each year (in particular we referred to the April release of GO for each year). We compared Harispe IC wrt to Sanchez IC and Sanchez Adapted with respect to Zhou IC since first two IC are normalized. Complete results are reported in the appendix and  at the page \url{https://sites.google.com/site/evolutionofic/home/statistics-by-year}. For each year we used the Wilcoxon Sum Rank test to reject the null Hypotesis $H_0$: \textit{there is no difference among distribution}. Results confirmed with a p-value less than 0,05 that all the compared distribution in each year were different. Thus we may conclude that the use of a different IC may yield to different conclusion. Finally we may affirm that even the choice of the IC formulation should be carefully considered since it is a possible cause of bias.

\section{\bf Conclusion}
\label{sec:conclusion}
 The GO and its annotations to gene products are now an integral part of functional analysis. Recently, the evaluation of similarity among gene products starting from their annotations (also referred to as semantic similarities) has become an increasing area in bioinformatics. While many research on updates to the structure of GO and on the annotation corpora have been made, the impact of GO evolution on semantic similarities is quite unobserved.  Here we extensively analyze how GO changes that should be carefully considered by all users of semantic similarities. GO changes in particular have a big impact on information content (IC) of GO terms. Since many semantic similarities rely on calculation of IC it is obvious that the study of these changes should be deeply investigated.  Here we consider GO versions from 2005 to 2014 and we calculate IC of all GO Terms considering five different formulation. Then we compare these results. Analysis confirm that there exists a statistically significant difference among different calculation on the same version of the ontology (and this is quite obvious) and there exists a statistically difference among the results obtained with different GO version on the same IC formula. Results evidence there exist a remarkable bias due to the GO evolution that has not been considered so far. Possible future works should keep into account this consideration.

\section*{\bf Acknowledgments}
This work has been partially founded by project PON Smartcities DICET-INMOTO-ORCHESTRA PON04a2\_D. \\Authors thank Dr. Sebastien Harispe for suggestions using Semantic Library Toolkit.

\section{Appendix}
\subsection{Statistics by Year}
The appendix reports the main descriptive statistics for the IC formulation year by year. For each year we used the GO version released on April.
The formulation of IC are those provided by Zhou et al \cite{zhou2008new}, Sanchez et al \cite{sanchez2011ontology}, Sanchez modified by Harispe \cite{Pesquita2009}, and Harispe et al \cite{harispe2013framework}.

\begin{table}[ht]	
\centering
\label{tab:stat2006}
\caption{Comparison of IC distribution - Year 2006}
\begin{tabular}{|c|c|c|c|c|}			\hline					
Value - Measure	&	Zhou	&	Sanchez Ad.	&	Sanchez	&	Harispe	\\ \hline
Mean	&	0,826	&	0,894	&	9,247	&	11,214	\\ \hline
SD	&	0,090	&	0,079	&	0,512	&	1,136	\\ \hline
Median	&	0,851	&	0,926	&	9,426	&	11,372	\\ \hline
1st Quartile	&	0,790	&	0,883	&	9,293	&	10,930	\\ \hline
2nd Quartile	&	0,851	&	0,926	&	9,426	&	11,372	\\ \hline
3th Quartile	&	0,875	&	0,926	&	9,426	&	11,824	\\ \hline
\end{tabular}									
\end{table}									

\begin{table}[ht]	
\centering
\label{tab:stat2007}
\caption{Comparison of IC distribution - Year 2007}
\begin{tabular}{|c|c|c|c|c|}	
\hline
Statistics - IC	&	Zhou	&	Sanchez Ad.	&	Sanchez	&	Harispe	\\ \hline
Mean	&	0,834	&	0,896	&	9,434	&	11,434	\\ \hline
SD	&	0,077	&	0,077	&	0,494	&	1,088	\\ \hline
Median	&	0,851	&	0,928	&	9,606	&	11,552	\\ \hline
1st Quartile	&	0,796	&	0,886	&	9,463	&	11,216	\\ \hline
2nd Quartile	&	0,851	&	0,928	&	9,606	&	11,552	\\ \hline
3th Quartile	&	0,875	&	0,928	&	9,606	&	12,004	\\ \hline
\end{tabular}
\end{table}

\begin{table}[ht]	
\centering
\label{tab:stat2008}
\caption{Comparison of IC distribution - Year 2008}
\begin{tabular}{|c|c|c|c|c|}	
\hline
Statistics - IC	&	Zhou	&	Sanchez Ad.	&	Sanchez	&	Harispe	\\ \hline
Mean	&	0,837	&	0,896	&	9,515	&	11,539	\\ \hline
SD	&	0,077	&	0,078	&	0,498	&	1,102	\\ \hline
Median	&	0,851	&	0,928	&	9,689	&	11,704	\\ \hline
1st Quartile	&	0,807	&	0,887	&	9,555	&	11,298	\\ \hline
2nd Quartile	&	0,851	&	0,928	&	9,689	&	11,704	\\ \hline
3th Quartile	&	0,881	&	0,928	&	9,689	&	12,087	\\ \hline
\end{tabular}
\end{table}

\begin{table}[ht]	
\centering
\label{tab:stat2009}
\caption{Comparison of IC distribution - Year 2009}
\begin{tabular}{|c|c|c|c|c|}	
\hline
Statistics - IC	&	Zhou	&	Sanchez Ad.	&	Sanchez	&	Harispe	\\ \hline
Mean	&	0,840	&	0,896	&	9,586	&	11,647	\\ \hline
SD	&	0,078	&	0,079	&	0,505	&	1,136	\\ \hline
Median	&	0,851	&	0,929	&	9,761	&	11,840	\\ \hline
1st Quartile	&	0,808	&	0,887	&	9,627	&	11,370	\\ \hline
2nd Quartile	&	0,851	&	0,929	&	9,761	&	11,840	\\ \hline
3th Quartile	&	0,896	&	0,929	&	9,761	&	12,246	\\ \hline
\end{tabular}
\end{table}

\begin{table}[ht]	
\centering
\label{tab:stat2010}
\caption{Comparison of IC distribution - Year 2010}
\begin{tabular}{|c|c|c|c|c|}	
\hline
Statistics - IC	&	Zhou	&	Sanchez Ad.	&	Sanchez	&	Harispe	\\ \hline
Mean	&	0,840	&	0,896	&	9,745	&	11,772	\\ \hline
SD	&	0,078	&	0,079	&	0,512	&	1,156	\\ \hline
Median	&	0,851	&	0,930	&	9,921	&	11,939	\\ \hline
1st Quartile	&	0,817	&	0,889	&	9,788	&	11,595	\\ \hline
2nd Quartile	&	0,851	&	0,930	&	9,921	&	11,939	\\ \hline
3th Quartile	&	0,896	&	0,930	&	9,921	&	12,345	\\ \hline
\end{tabular}
\end{table}

\begin{table}[ht]	
\centering
\label{tab:stat2011}
\caption{Comparison of IC distribution - Year 2011}
\begin{tabular}{|c|c|c|c|c|}	
\hline
Statistics - IC	&	Zhou	&	Sanchez Ad.	&	Sanchez	&	Harispe	\\ \hline
Mean	&	0,835	&	0,896	&	9,745	&	11,856	\\ \hline
SD	&	0,078	&	0,079	&	0,512	&	1,167	\\ \hline
Median	&	0,843	&	0,930	&	9,921	&	12,001	\\ \hline
1st Quartile	&	0,810	&	0,889	&	9,788	&	11,682	\\ \hline
2nd Quartile	&	0,843	&	0,930	&	9,921	&	12,001	\\ \hline
3th Quartile	&	0,888	&	0,930	&	9,921	&	12,486	\\ \hline
\end{tabular}
\end{table}

\begin{table}[ht]	
\centering
\label{tab:stat2012}
\caption{Comparison of IC distribution - Year 2012}
\begin{tabular}{|c|c|c|c|c|}	
\hline
Statistics - IC	&	Zhou	&	Sanchez Ad.	&	Sanchez	&	Harispe	\\ \hline
Mean	&	0,836	&	0,897	&	9,812	&	11,935	\\ \hline
SD	&	0,078	&	0,079	&	0,511	&	1,170	\\ \hline
Median	&	0,843	&	0,931	&	9,986	&	12,066	\\ \hline
1st Quartile	&	0,810	&	0,890	&	9,853	&	11,778	\\ \hline
2nd Quartile	&	0,843	&	0,931	&	9,986	&	12,066	\\ \hline
3th Quartile	&	0,888	&	0,931	&	9,986	&	12,551	\\ \hline
\end{tabular}
\end{table}

\begin{table}[ht]	
\centering
\label{tab:stat2013}
\caption{Comparison of IC distribution - Year 2013}
\begin{tabular}{|c|c|c|c|c|}	
\hline
Statistics - IC	&	Zhou	&	Sanchez Ad.	&	Sanchez	&	Harispe	\\ \hline
Mean	&	0,834	&	0,897	&	9,882	&	12,097	\\ \hline
SD	&	0,078	&	0,081	&	0,518	&	1,204	\\ \hline
Median	&	0,837	&	0,931	&	10,053	&	12,250	\\ \hline
1st Quartile	&	0,808	&	0,891	&	9,935	&	11,844	\\ \hline
2nd Quartile	&	0,837	&	0,931	&	10,053	&	12,250	\\ \hline
3th Quartile	&	0,880	&	0,931	&	10,053	&	12,761	\\ \hline
\end{tabular}
\end{table}

\begin{table}[ht]	
\centering
\label{tab:stat2014}
\caption{Comparison of IC distribution - Year 2014}
\begin{tabular}{|c|c|c|c|c|}	
\hline
Statistics - IC	&	Zhou	&	Sanchez Ad.	&	Sanchez	&	Harispe	\\ \hline
Mean	&	0,834	&	0,897	&	9,914	&	12,143	\\ \hline
SD	&	0,078	&	0,081	&	0,519	&	1,214	\\ \hline
Median	&	0,837	&	0,931	&	10,086	&	12,283	\\ \hline
1st Quartile	&	0,808	&	0,891	&	9,968	&	11,877	\\ \hline
2nd Quartile	&	0,837	&	0,931	&	10,086	&	12,283	\\ \hline
3th Quartile	&	0,880	&	0,931	&	10,086	&	12,858	\\ \hline
\end{tabular}
\end{table}

\end{document}